\documentclass[preprint2,aas2pp4,natbib209]{aastex}
\usepackage{graphicx}
\usepackage[colorlinks=true,citecolor=blue]{hyperref}
\bibpunct{(}{)}{;}{a}{}{,}
\usepackage[switch,pagewise]{lineno}
\usepackage{subfigure}
\usepackage{overpic}
\usepackage{multirow}
\usepackage{ulem}
\usepackage{color}
\usepackage{amsmath}
\usepackage{txfonts}
\usepackage{cases}
\usepackage{natbib}

\shorttitle{type IIIb structure observing with LOFAR}
\shortauthors{X.Y.Chen et al.}

\begin{document}
\title{Fine Structures of Solar Radio Type III Bursts and their Possible Relationship with Coronal Density Turbulence}
\author{Xingyao Chen$^{1,2,3}$, Eduard P. Kontar$^{3}$, Sijie Yu$^4$, Yihua Yan$^{1,2}$, Jing Huang$^{1,2}$, Baolin Tan$^{1,2}$}
\affil{$^1$Key Laboratory of Solar Activity, National Astronomical Observatories Chinese Academy of Sciences, Beijing 100012, China\\
	$^2$University of Chinese Academy of Sciences, Beijing 100049, China\\
	$^3$School of Physics and Astronomy, University of Glasgow, Glasgow G12 8QQ, UK\\
	$^4$New Jersey Institute of Technology, 323 Martin Luther King Boulevard, Newark, NJ 07102, USA}
\email{xychen@nao.cas.cn}

\begin{abstract}
Solar radio type III bursts are believed to be the most sensitive signature of near-relativistic electron beam propagation in the corona. A solar radio type IIIb-III pair burst with fine frequency structures, observed by the Low Frequency Array (LOFAR) with high temporal ($\sim10$ ms) and spectral (12.5 kHz) resolutions at 30 - 80 MHz, is presented. The observations show that the type III burst consists of many striae, which have a frequency scale of about 0.1 MHz in both the fundamental (plasma) and the harmonic (double plasma) emission. We investigate the effects of background density fluctuations based on the observation of striae structure to estimate the density perturbation in solar corona. It is found that the spectral index of the density
fluctuation spectrum is about $-1.7$, and the characteristic spatial scale of the density perturbation is around $700$~km. This spectral index is very close to a Kolmogorov turbulence spectral index of $-5/3$, consistent with a turbulent cascade. This fact indicates that the coronal turbulence may play the important role of modulating the time structures of solar radio type III bursts, and the fine structure of radio type III bursts could provide a useful and unique tool to diagnose the turbulence in the solar corona.
\end{abstract}


\section{Introduction}
\label{sec01}

Solar radio type III bursts are a common signature of near-relativistic electrons streaming through plasma
in the solar corona and interplanetary space,
and they provide a useful way to remotely trace these electrons \citep[e.g.][]{1950AuSRA...3..541W,2008A&ARv..16....1P}.
Type III bursts usually drift from high to low frequencies with a fast drifting rate corresponding to an exciter speed of $\approx c/3$,
where $c$ is the speed of light.
They are believed to be generated by outward propagating beams of non-thermal electrons along open magnetic field lines. \cite{1958SvA.....2..653G} proposed plasma emission
as the main generation mechanism of solar radio type III bursts, which is the most widely accepted model.
As the electrons propagate along magnetic field lines,
the faster ones arrive at a remote location faster,
forming an unstable distribution leading to a bump-on-tail instability.
This instability generates Langmuir waves at the local plasma frequency,
$f_{p}$. The coalescence (or decay) of Langmuir waves and low frequency ion-sound waves may produce radiation near the electron
plasma frequency $f_{p}$, which is called the fundamental emission.
The coalescence of Langmuir waves and scattered Langmuir waves
may produce the second harmonic emission $2f_{pe}$.
This theory has been discussed and refined by many authors \citep[e.g.][]{1964NASSP..50..357S,1985ARA&A..23..169D,1987SoPh..111...89M,1994ApJ...422..870R,1998ARA&A..36..131B,2014A&A...572A.111R,2016A&A...586A..19T,2017SoPh..292..117L}.

The fundamental and harmonic components of solar radio type III bursts are frequently observed by broadband dynamic spectrometers. They sometimes show short-duration and narrow-band intermittent
features in both time and frequency.
The so-called solar radio type IIIb bursts were identified by \cite{1972A&A....20...55D} as a chain of several elementary bursts,
which appear in the dynamic spectrum as either single, double, or triple narrow-banded striations \citep{1975STIN...7618028S}.
These narrowband features are known as striae,
and a burst composed of such features is known as type IIIb burst.
These striae structures have a very narrow bandwidth of
$\Delta f/f \simeq 1\%$ \citep[e.g.][]{2010AIPC.1206..445M,2017Mugundhan}.
The observations of the fundamental-harmonic structure
in Type IIIb-III pairs \citep[e.g.][]{1984SoPh...91..377A}
further supports the plasma emission mechanism of these bursts.
However, there are several open questions.
\citet{1976ApJ...207..605S} suggest that the strong beam-plasma interaction leads to Langmuir wave amplification with subsequent Langmuir wave decay into transverse radio waves.
In such a scenario, a chain of striae is produced
through the modulational instability.
However, \cite{1998ApJ...509..471C} argue that the predicted wavenumbers and bandwidths of Langmuir waves are too large for modulational instability to occur directly.
According to \cite{1975SoPh...40..421T},
type IIIb bursts are generated by electron beams, which propagate through nonuniform plasma. They postulate that locally
overdense and/or underdense regions in the plasma increase or decrease the interaction lengths of the beam for different frequency ranges, resulting in striae.
\cite{2001A&A...375..629K} numerically show that if fast electron beams propagate in the solar corona with density fluctuations, the Langmuir turbulence would be spatially nonuniform with regions of enhanced and reduced levels of Langmuir waves,
which can produce the chains of striae.
\citet{2001A&A...375..629K} and \citet{2010ApJ...721..864R} show that the background solar wind electron density fluctuations cause plasma waves to be non-uniform in space, with larger amplitudes
and smaller length scales of density fluctuations having the largest effect.
These modulations of Langmuir waves can lead to the modulation of subsequent plasma radio emission.
The numerical investigations of \cite{2012SoPh..279..173L,2015ApJ...807...38V} using the effects of density irregularities also support this idea.
\cite{2012SoPh..279..173L}, also using numerical methods, have further confirmed that enhanced density structures along the beam path and either electron or ion temperature enhancements of unknown origin can reproduce striae-like features.
However, they find that the second-harmonic emission should
be stronger than the fundamental emission,
which is inconsistent with the observations.
\cite{2013ApJ...779...31Z} investigate the effects of homochromous Alfv{\'e}n waves on electron-cyclotron maser emission proposed by \cite{2002ApJ...575.1094W}, that may be responsible for producing type III bursts. They show that the growth rate of the O-mode wave will be significantly modulated by homochromous Alfv{\'e}n waves, which could produce type IIIb radio bursts.

In summary, it is widely believed that in the plasma emission mechanism, density inhomogeneities in the background plasma could create a clumpy distribution of Langmuir waves and generate the type IIIb fine structure \citep[e.g.][and so on]{1975SoPh...40..421T,1983SoPh...87..359M}.
The density fluctuation spectrum in the interstellar medium and solar wind often reveals a Kolmogorov-like scaling with a spectral slope of $-5/3$
in wavenumber space \citep[e.g.][and so on]{1941DoSSR..30..301K,1995ARA&A..33..283G,2004ARA&A..42..211E,2010MNRAS.402..362S}. However, the physical processes leading to a Kolmogorov-like turbulent density fluctuation spectrum are not yet fully understood.
Here, our observations of a type IIIb radio burst support the result that density turbulence in the corona is likely to cause the striae structure, and that the escaping plasma radio emission fluctuates
along the direction that the beam propagates with a power-law flux fluctuation spectrum.

Because of the modest sensitivity of previous solar radio telescopes,
there are few studies of fine structures of type III bursts.
Observations with the LOw Frequency ARray \citep[LOFAR;][]{2013A&A...556A...2V} now permit us to analyze type-III/IIIb emission with high temporal and spectral resolutions,
which were unavailable before.
Here, we use LOFAR data to study the fine structures of type III radio bursts observed in the fundamental and harmonic components
at 30-80 MHz.
We aim to investigate the effects of background density fluctuations based on the observations of those striae structure and estimate the properties of the density perturbation in the solar corona.
The article is arranged as follows. Section 2 describes the details of the observations. Section 3 presents the discussions and conclusions.

\section{Observations}
\label{sec02}

We use spectral and imaging data from LOFAR, a large radio telescope located mainly in the Netherlands and completed in 2012 by the Netherlands Institute for Radio Astronomy (ASTRON). LOFAR can make observations in the 10 - 240 MHz frequency range with two types of antenna: the Low Band Antenna (LBA) and the High Band Antenna (HBA), optimized for 30 - 80 MHz and 120 - 240 MHz, respectively.
The antennas are arranged in clusters (stations) that are spread out over an area of more than 1000 km in diameter,
mainly in the Netherlands and partly in other European countries.
The LOFAR stations in the Netherlands reach baselines of about 100 km. LOFAR currently receives data from 46 stations, including 24 core stations, 14 remote stations in the Netherlands, and 8 international stations.
Each core and remote station has 48 HBA and 96 LBA  antennas with a total of 48 digital Receiver Units.
One of the main LOFAR projects is to study solar physics and space weather. Currently, there are only few analyses of solar observations with LOFAR, such as the study of solar radio type III-like bursts \citep{2014A&A...568A..67M,2017A&A...606A.141R}.
In this work, we investigate a solar radio type III burst with fine structures using 24 LOFAR core stations, including the broadband spectrum
and imaging observations at frequencies of 30 - 80 MHz
from the LBA array. These 24 core stations share
the same clock and have a maximum baseline of $\sim 3.5$~km \citep{2013A&A...556A...2V}. The tied-array beam forming mode
simultaneously provides a frequency resolution of 12.5 kHz
with a high time cadence of $\sim10$~ms.

We also use observational data at ultraviolet (UV)
and extreme ultraviolet (EUV) wavelengths taken
from the Atmospheric Imaging Assembly onboard the Solar Dynamics Observatory \citep[AIA/SDO;][]{2012SoPh..275...17L}.
Magnetic field data is collected from the Helioseismic Magnetic Imager onboard SDO \citep[HMI/SDO;][]{2012ASPC..462..352S}.
The pixel sizes of AIA and HMI are around $0.6''$,
with time cadences of 12 s and 45 s, respectively.

\subsection{Observation and data analysis}
\label{sec0201}
\begin{figure*}[ht!]
\includegraphics[width=0.88\linewidth]{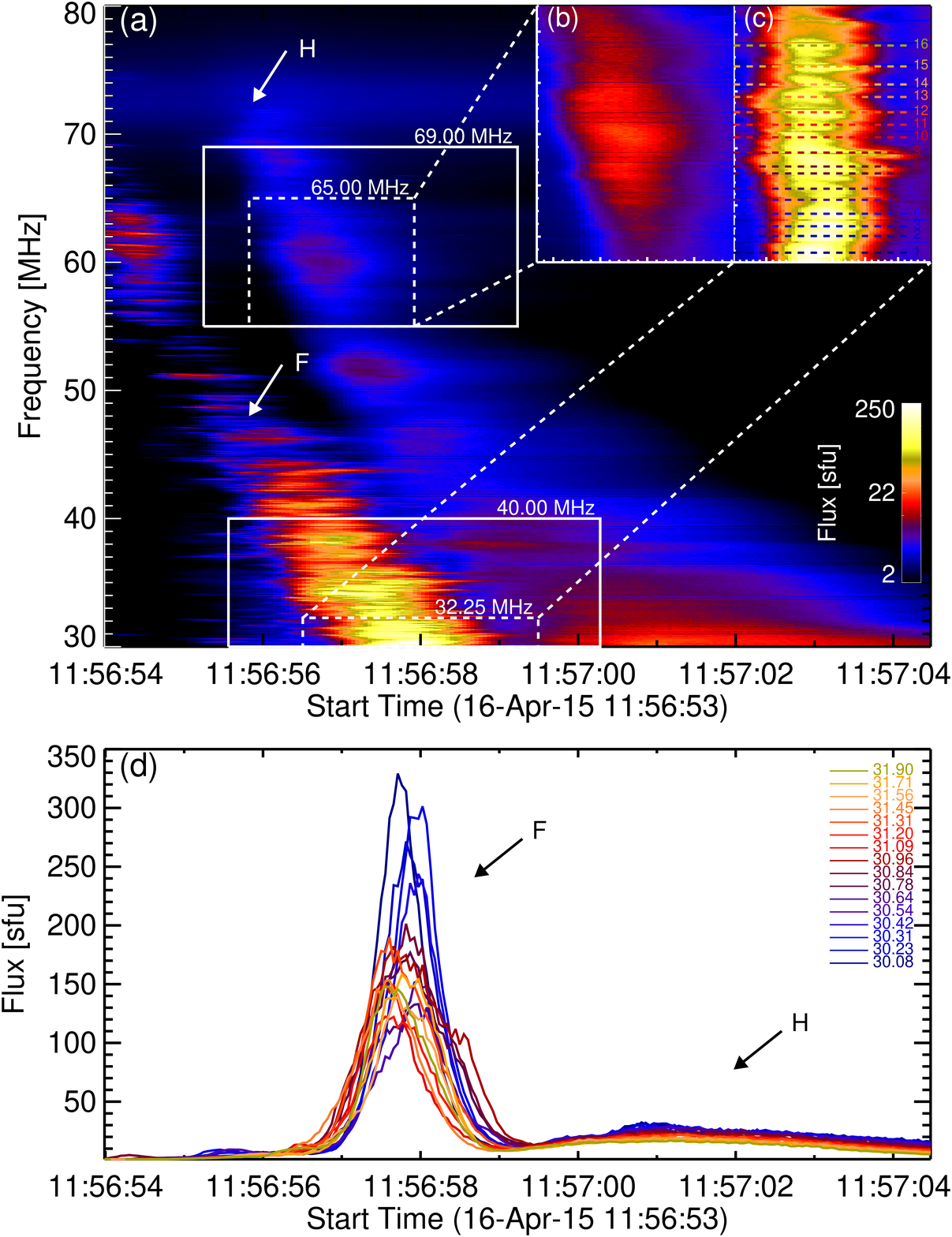}
\caption{(a) The dynamic spectrum of a type IIIb-type III burst pair in the frequency range of 30 - 80 MHz observed by LOFAR on 16 April 2015. F and H indicate the fundamental and second harmonic branches, respectively.
The solid rectangular boxes show the time and frequency ranges investigated in the analysis.
(b) An enlarged version of the harmonic components
from 55 to 65 MHz.
(c) The detailed spectrum between 30 and 32.25 MHz.
The dashed horizontal lines represent the different frequencies in Figure \ref{fig-1}(d).
(d) The fluxes of the F and H components selected from the striae structure in (c).
\label{fig-1}}
\end{figure*}

The type IIIb burst occurred at about 11:56:58 UT on 16 April 2015, between an impulsive C2.3 flare (starting at 11:23 UT and ending at 11:30 UT), and a long-duration C2.0 flare (starting at 12:23 UT and ending at 14:00 UT) in the active region NOAA12321.
{\color{black}The imaging of this event was analyzed by \cite{2017arXiv170806505K}. They compare the centroid positions,
sizes and areal extents of the fundamental and harmonic sources and they demonstrate that the propagation-scattering effects dominate the observed spatial characteristics of the radio burst images. Here, we examine the fine structures in the spectrum.}
Figure \ref{fig-1} presents the spectrum of the type III burst at a frequency range of 30 - 80 MHz observed by LOFAR. The type III burst is composed of two branches. One is the fundamental (F) branch occurring in the frequency range of 30 - 65 MHz, and the other is the second harmonic (H) branch, which is much weaker than the fundamental, occurring in the frequency range of 30 - 80 MHz. The mean ratio between the H and F branches  {\color{black}is about 1.6 for a given time.  This is consistent with other observational values of 1.5 -1.8 \citep{1985srph.book..289S}.
The frequency drift rate of the type III burst is about $-7.3$ and $-13.8$ MHz s$^{-1}$ for the F and H components, respectively.}
Here, a negative frequency drifting rate implies that the related electron beams propagated from low to high altitudes in the solar corona. The dynamic spectrum shows that there are distinct modulated fine structures in the fundamental component. The harmonic component also shows such structures but they are much weaker.

\begin{figure*}[ht!]
\includegraphics[width=\linewidth]{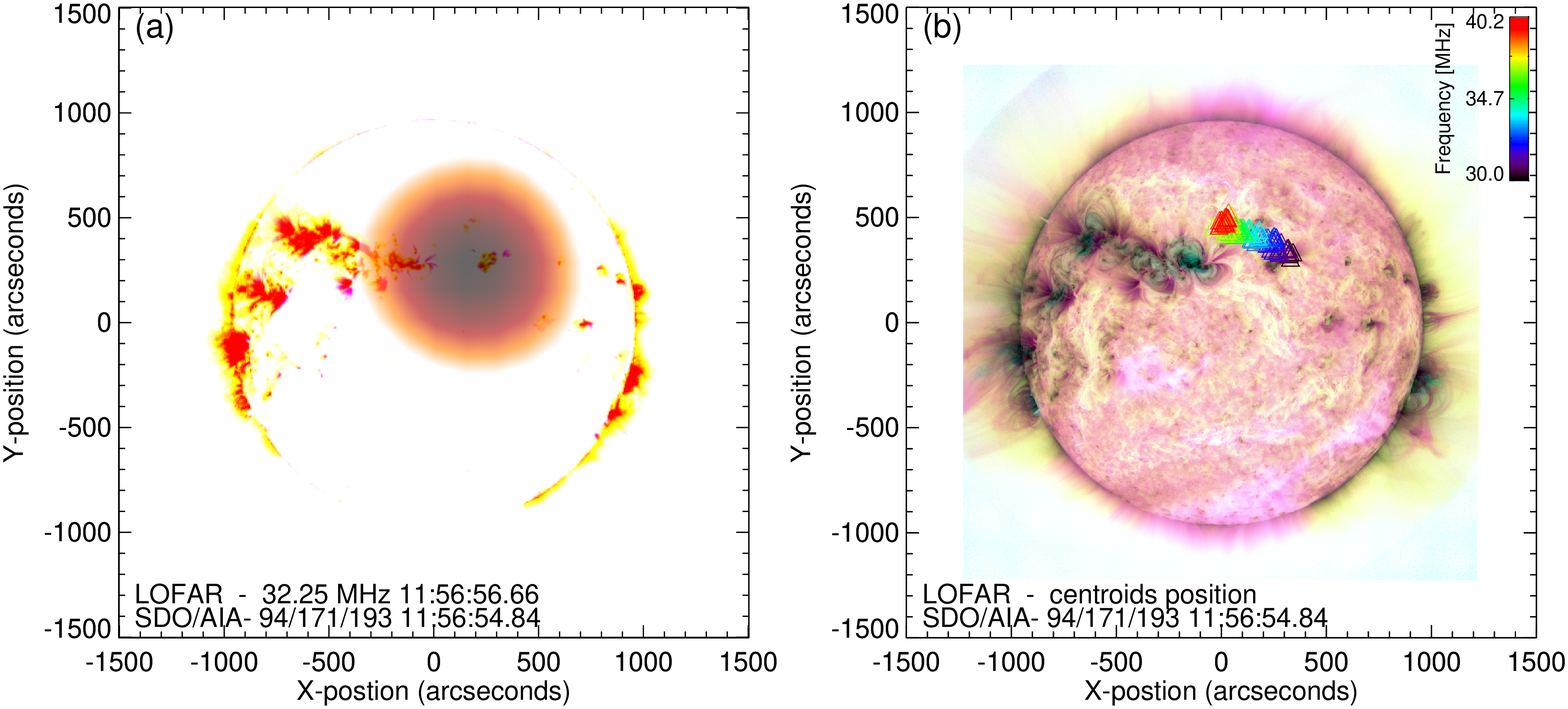}
\caption{(a) A composite of AIA images: 94 $\r{A}$ (red),171 $\r{A}$ (green), and 193 $\r{A}$ (blue) {\color{black} highlighting the active regions (most of the red background)}.
A 32.25 MHz LOFAR source at 11:56:56 UT is overlaid. (b) A composite image {\color{black}of AIA at the same time as (a), but using a different color table.} The triangle symbols indicate the centroids of the LOFAR source in frequency and time, corresponding to the green asterisks in Figure \ref{fig-3}(a).
\label{fig-2}}
\end{figure*}

{\color{black} Figure \ref{fig-2} presents the imaging observations of this solar radio type III source obtained by LOFAR.
A semi-transparent ellipse in Figure \ref{fig-2}(a) is the LOFAR source image at a frequency of 32.25 MHz and a time of 11:56:56 UT.
This is overlaid on a composite AIA image
displaying the 94 \AA, 171 \AA, and 193 \AA~channels.
The radio source is slightly larger than the intrinsic source size
and the harmonic (H) component source is larger than the fundamental (F) emission component.
\citet{2017arXiv170806505K} explain that both F and H radio sources are enlarged mainly due to propagation-scattering effects from the source in the corona.}
{\color{black}At the same time, we get the centroid positions
by fitting the sources with a two-dimensional gaussian at each frequency channel in the range of $30-40$~MHz.
The found centroid positions are shown as triangles in Figure \ref{fig-2}(b).} We find that the centroids of higher frequency sources are located more closely to the active region than the centroids of lower frequency sources.
We utilize the Potential Field Source Surface \citep[PFSS;][]{2003SoPh..212..165S} model to extrapolate the large-scale magnetic field topology of the region.
The results show that the locations of radio sources tend to propagate along the open magnetic field lines of a nearby active region.
{\color{black} By scrutinizing the radio emission at the same time,
we find that both the fundamental and harmonic branches have fine structure. In the dynamic spectrum,
the fundamental branch (Figure \ref{fig-1}(a).)
is composed of about 70 striae.
Each stria has a frequency bandwidth of about $\sim100$ kHz
and a lifetime of about 1 s, consistent with other observations of short ($\sim1.2$ s) and narrow-band ($\sim100$ kHz) stria bursts \citep{1972A&A....20...55D,1974SoPh...39..223B,2010AIPC.1206..445M}.}
We zoom-in on the spectrum in Figure \ref{fig-1}(c) and the fundamental emission shows 16 striae in the frequency range from 30 MHz to 32 MHz. Some of the striae overlap with each other.
Figure \ref{fig-1}(a) also shows that the brightness of striae decrease with frequency (see the profiles of Figure \ref{fig-1}(d) in detail).
The second harmonic branch has a relatively faint structure (Figure \ref{fig-1}(a)) compared to the fundamental. A small part of the second harmonic dynamic spectra is displayed in Figure \ref{fig-1}(b). We find that the second harmonic branch is also composed of many small striae. The tails in the F and H radiation are also visible in the dynamic spectrum.

\begin{figure*}[ht!]
\includegraphics[width=\linewidth]{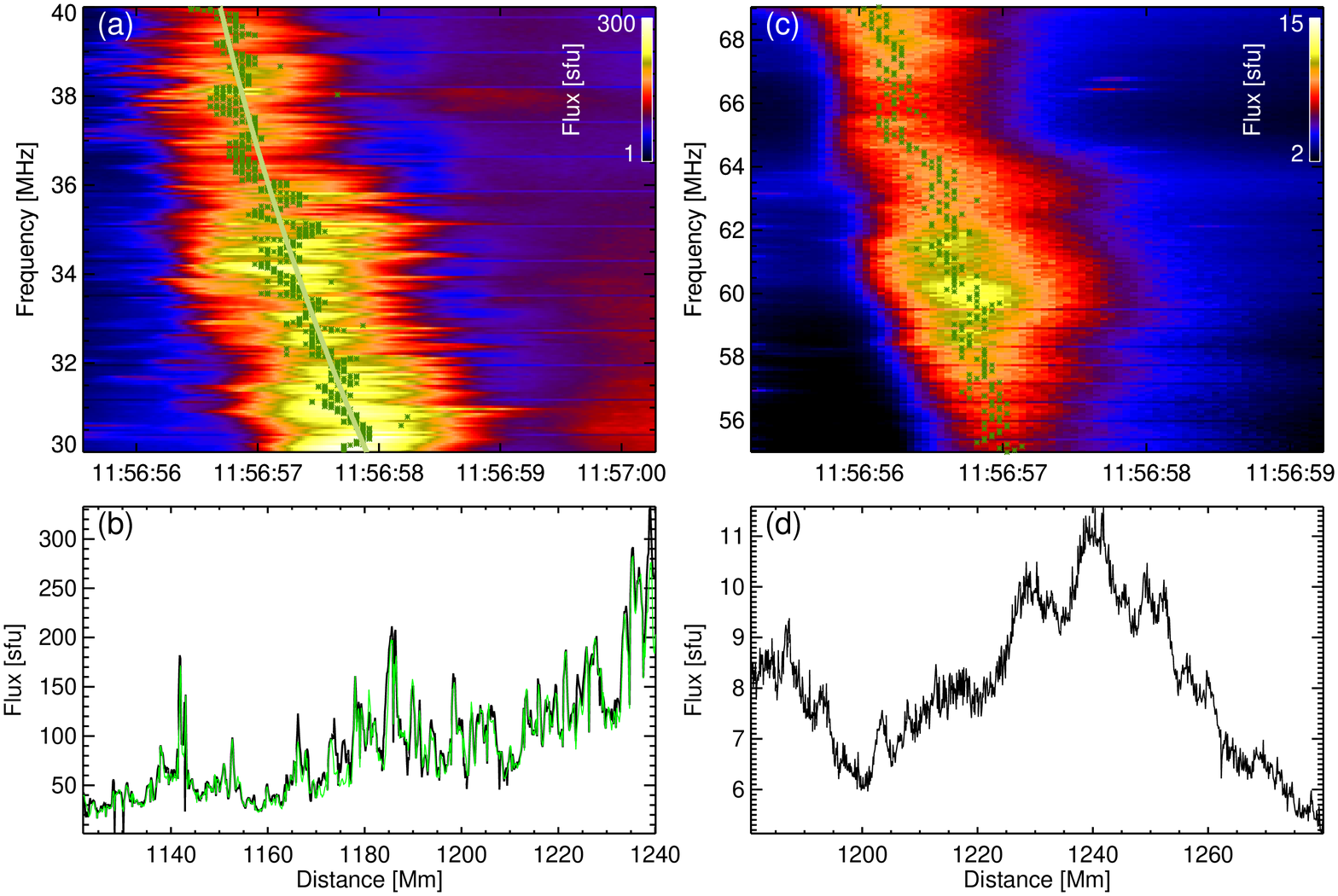}
\caption{\label{fig-3}(a) The spectrum of the fundamental (F) component at 30-40 MHz from 11:56:55 UT to 11:57:00 UT. The green asterisks show the positions of the fitted gaussian peak times for the F component at each frequency. The green line represents the best fit through all the positions of the fitted gaussian peaks using a least-square polynomial fitting function.
(b) A plot of flux versus distance using a Newkirk coronal density model (black solid line). The green line indicates the flux of the fitted gaussian peak times for the F component at each frequency (green asterisks (a)).
(c) The spectrum of the harmonic (H) component at 55-69 MHz. The green asterisks again show the fitted gaussian peak times of the H components at each frequency.
(d) Radio flux versus distance as in (b), but for the H components.}
\end{figure*}

\begin{figure*}[ht!]
\includegraphics[width=\linewidth]{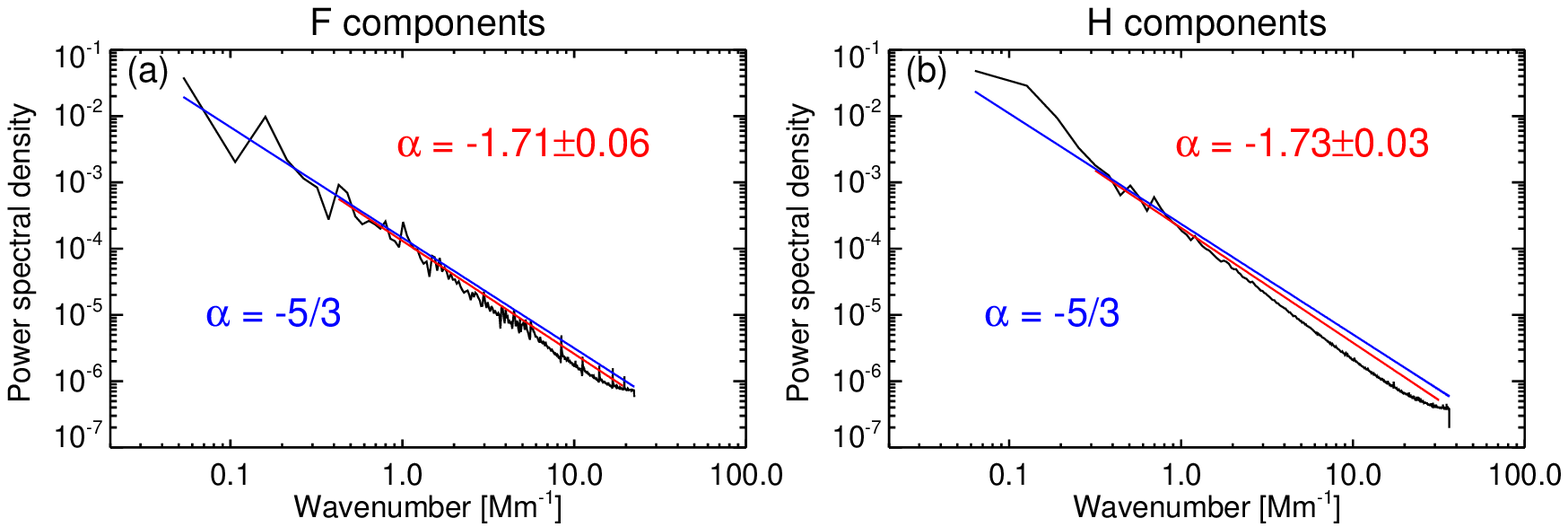}
\caption{\label{fig-4}The power spectral density of flux fluctuations (black curve).
(a) The Fourier transform of the autocorrection function of fundamental flux at 30-40 MHz, see Figure \ref{fig-3}(b).
(b) The Fourier transform of the autocorrection function of the harmonic (H) component flux in Figure \ref{fig-3}(d).
The blue line is the Kolmogorov function of $P(k)\sim k^{-5/3}$ line.
The red lines are obtained from least-square fitting of the power spectral density with the form $P(k)\sim k^{-\alpha}$ .}
\end{figure*}

The dynamic spectrum of the fundamental emission in the frequency range of 30 - 40 MHz during 11:56:55 - 11:57:00 UT 
is plotted in Figure \ref{fig-3}(a).
Here, bright green asterisks mark the positions of peak times obtained from the gaussian fits, for the F component at each frequency. The green line is obtained by fitting all the peak times using a least-square polynomial fitting method.
{\color{black}There are several coronal density models used to estimate the distance of the source corresponding to an emitting frequency \citep{1961ApJ...133..983N,1977SoPh...55..121S,1998SoPh..183..165L,1999A&A...348..614M,2009ApJ...706L.265C}.}
{\color{black}In our analysis, the precise value of the density or the distance weakly affects the results.
From the following analysis, the different distance intervals seem to have little impact on the flux curve, as we now demonstrate.}
We compare the distance of sources between 30 and 40 MHz using different density models. We find only a slight difference in the distance away from the solar center at different frequencies, and the distance intervals are around 100 Mm within the frequency range of 30 - 40 MHz, in the frame of different density models. The determined distance intervals for the F components for different models are $115.9$,~$100.2$,~$93.3$,~and $86.4$~Mm \citep{1961ApJ...133..983N,1977SoPh...55..121S,1998SoPh..183..165L,1999A&A...348..614M}.
As an example, we show the analysis 
using the Newkirk model (\ref{eq-2}), 
to estimate the distance:
\begin{equation}
\label{eq-2}
n(r)=n_0\times 10^{4.32/r},
\end{equation}
where $n_0$ is set to be $4.2\times{10^4}$~cm$^{-3}$. Then
the local plasma frequency is
\begin{equation}
\label{eq-1}
f_{pe}=\frac{1}{2\pi}\sqrt{\frac{4{\pi}e^2n(r)}{m}}.
\end{equation}
After the above transformation, the emission flux at different distances along the radiation propagating path is shown in Figure \ref{fig-3}(b).
Then, the radio emission originates
between the distances of $1121$ and $1240$~Mm ($1.6-1.77R_\odot$) from the solar center, where $R_\odot$ is the solar radius.
The flux as a function of distance shows large fluctuations.
The flux is also growing with distance, and the larger flux 
values are observed at lower frequencies.

\subsection{The power spectrum of radio flux fluctuations}
{\color{black} By converting frequency into distance using
Newkirk's density model (Equations \ref{eq-2}, \ref{eq-1}),
a flux-distance relation (see Figure \ref{fig-3})
is obtained for the analysis of flux fluctuations.
According to the Wiener-Khinchin theorem, the power spectrum density function is the Fourier transform of the autocorrection function.
So, the power spectrum of the flux fluctuation
in the frequency domain is a function of $k$,
which is the wavenumber (cycles per Mm) given by}
\begin{equation}
\label{eq-3}
P(k)=\int_{-\infty}^{\infty} R(\lambda) e^{-ik\lambda}d\lambda ,
\end{equation}
where $P(k)$ is the Fourier transform of the autocorrelation function
given by $R(\lambda) = \langle F(r)F(r+\lambda)\rangle $,
and where the angular brackets $\langle ...\rangle$ denote an ensemble average.
$r$ is the radial distance from the solar center and $F(r)$ is the radio flux observed. The exciter of the radio burst is assumed to be radially propagating.
We make the Fourier transform of $R(\lambda)$ and obtain the power spectrum density of the radio flux $F(r)$ as shown in Figure \ref{fig-4}, where the abscise axis shows wavenumber $k$.
{\color{black} The blue line in the figure is the theoretical
Kolmogorov function of $P(k)\sim k^{-5/3}$
with a spectral index of $-5/3$, which is plotted for comparison.
The red line is a best fit to the spectrum power in wavenumber space.
It shows that the spectral index is $-1.71\pm 0.06$,
which is close to the Kolmogorov spectral index of $-1.67$.
We take into account the flux $F(x)$ uncertainties, which are typically around $1$~s.f.u. for the background flux level before
the burst for each frequency \citep{2017arXiv170806505K}.
We calculate the uncertainties from error propagation
during the spectral analysis.
{\color{black} We also consider the errors from the fit and take the 1-sigma uncertainty for the fitted spectral index.
 The calculated uncertainty is about $0.06$ for the spectral index. Those errors are shown in Figure \ref{fig-4}.}
Furthermore, we also use the fast Fourier analysis
on the emission flux of type IIIb striae and the corresponding least-squares fitting line, which shows an approximate result.
The same analysis has also been used for the second harmonic branch shown in Figure \ref{fig-4}(b),
which gives a similar spectral index of $-1.73\pm0.03$.
Using other density models, the spectral indexes for the fundamental (F) and harmonic (H) components are:
$-1.72\pm 0.05$ and $-1.75 \pm 0.02$ for Saito's density model, $-1.72\pm 0.05$ and $-1.77\pm 0.02$ for Leblanc's density model,
and $-1.73\pm0.05$ and $-1.78\pm$0.02 for Parker's density model.
We also try models that are 2 times and 5 times the Newkirk density profile and the calculated intervals are larger than the commonly used density models. For $2\times$ Newkirk's density profile,
the distance intervals for the fundamental and harmonic components
are $149.1$~Mm and $129.1$~Mm and the spectral indexes change to $-1.69\pm 0.07$ and $-1.71\pm0.03$ respectively.
For the 5$\times$ Newkirk's density profile, the distance intervals for the fundamental and harmonic components are $220.5$~Mm
and $194.8$~Mm and the spectral indexes are $-1.67 \pm 0.07$ and $-1.68 \pm 0.03$.
As shown by the above results, the different density models show similar results and all models provide spectral indices close to the Kolmogorov turbulence spectral index of $-5/3$. This result provides a plausible explanation for the effects of turbulence modulation on the F and H components along the radio radiation propagation. }

\begin{figure}[ht!]
\includegraphics[width=0.9\linewidth]{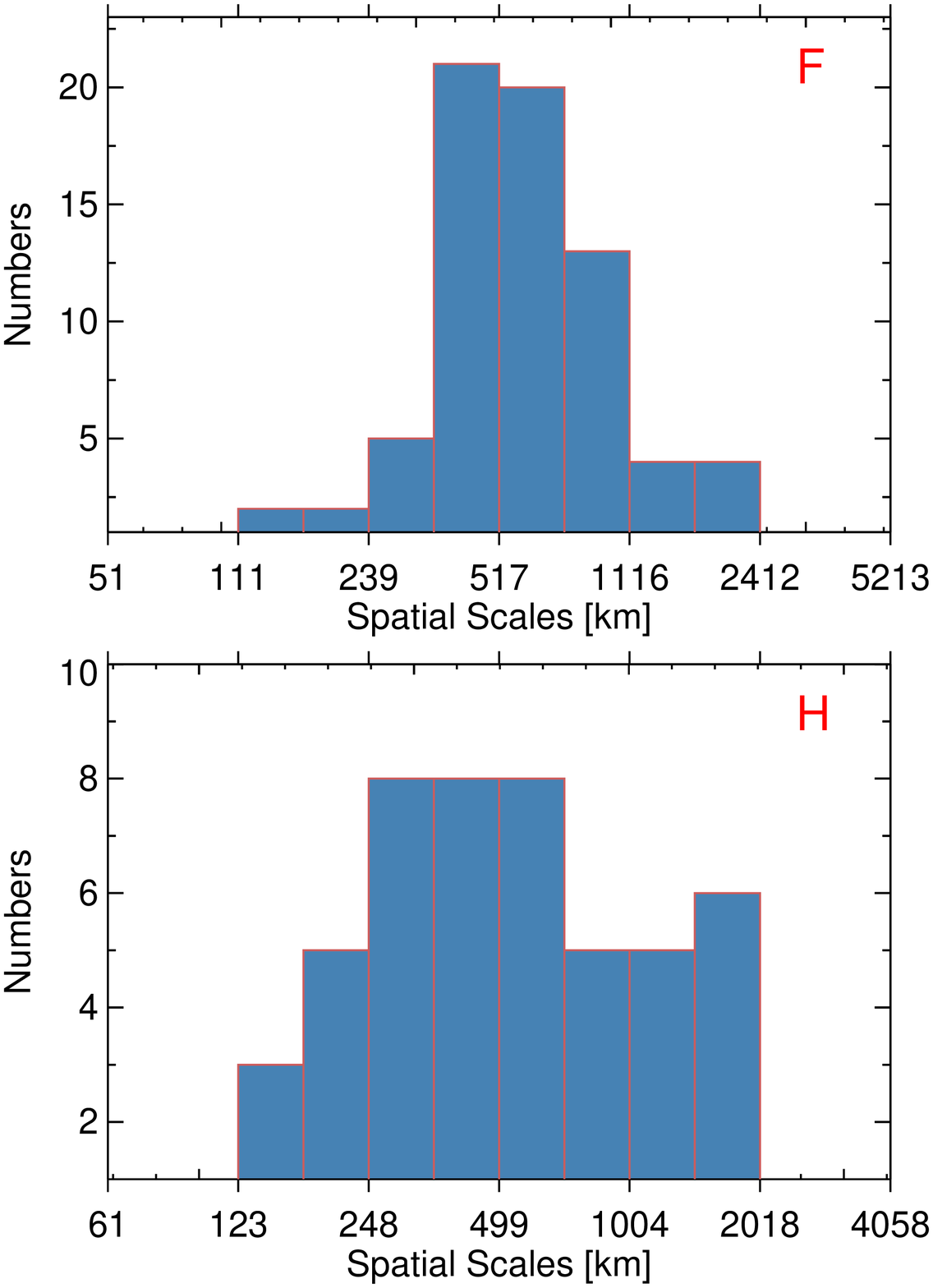}
\caption{\label{fig-5} Histogram of the spatial scales measured in both the fundamental (F) and harmonic (H) branches.}
\end{figure}

Under the interpretation of Type-IIIb burst turbulence modulation,
we estimate the spatial scale of turbulence elements using coronal density models.
The stria with a peak flux greater than $1\sigma$ is considered to be an individual striae. The full width at half maximum (FWHM) of selected spikes is defined as the frequency bandwidth of stria, which can be transformed into the spatial scale of a turbulence element.
The statistical histograms of spatial scales using the Newkirk density model are shown in Figure \ref{fig-5}. Here, we find that the average spatial scales of a turbulence element is about $700$~km, derived from the fundamental branch, and also $\sim 700$~km derived from the harmonic branch. These two values are consistent with each other and both the fundamental and harmonic show values distributed in a similar range from $\sim 100$~km to $\sim 2000$~km,
with a most probable spatial scale of around $700$~km.
This suggests that it is highly likely that the fundamental and harmonic branches of the type-IIIb radio bursts are modulated by the same turbulence.

\section{Discussion and conclusions}
\label{sec03}

{\color{black} Solar radio type III bursts are produced by near-relativistic electrons streaming through the corona and such bursts can provide important information about the background plasma properties.
The background plasma density corresponding to a frequency
of several tens of MHz in the high corona is nonuniform, and therefore, type III bursts should have flux variations associated with density fluctuations, as evident from numerical simulations \cite[e.g.][]{2001A&A...375..629K,2010ApJ...721..864R}.
Using high resolution LOFAR observations,
we find that the fundamental and second harmonic branches of the type IIIb-III pair burst,
within the frequency range of 30 - 80 MHz, have fine 0.1~MHz frequency structures, visible as a train of intermittent small striae.
Each distinct stria has a mean lifetime of about 1~s, with longer duration times at lower frequencies.
This is consistent with the empirical relation for the decay time of metric and kilometric type III bursts \citep{1992SoPh..141..335B}.
Under the turbulent density interpretation,
the flux variation with distance can provide a new method
of diagnosing density fluctuations.
The characteristic scales of radio flux fluctuations are distributed in the range from $\sim 100$~km to $\sim 2000$~km,
with the most probable spatial scale occurring around $700$~km.
The power spectrum of the radio emission fluctuations as a function of wavenumber is close to a power-law distribution, $P(k)\propto k^{\alpha}$, with spectral indices of $\alpha = -1.71 \pm 0.06$
and $\alpha = -1.73 \pm 0.03$ for the fundamental and harmonic components of type IIIb-type III pair, respectively.
These spectral indices are very close to a Kolmogorov density turbulence spectral index of $-5/3$, which is also consistent with near Earth in-situ observations. We also examined different density models, and they show different distance intervals. However, density models have little impact on the flux curve and they only slightly influence the determined spectral index. All the spectral indices determined from different density models are close to the Kolmogorov turbulence spectral index.} The evidence suggests that the escaping radio emission produced by near-relativistic electron beams  is modulated by the density turbulence present in the solar corona.

In this event, the fundamental - harmonic pair emission shows that the fine structures are better observed in the fundamental component.  However, the striae structure also appear in the harmonic component, although it is much weaker than in the fundamental component.
The intensity of the fundamental emission is 20-30 times higher
than that of the harmonic component, and the mean frequency band of each single striae is only 0.1~MHz (Figure \ref{fig-1}(d)).
The scattering off ions or the decay of Langmuir waves would produce the fundamental component, while the coalescence of two Langmuir waves would produce the harmonic component.
According to the plasma emission mechanism \cite[e.g.][]{1987SoPh..111...89M}, the fundamental and harmonic components should come from the same spatial location.
If an observation records harmonic radio radiation,
one possible reason for a lack of observable fine structure in the harmonic components might be due to a lack of sensitivity in the radio telescopes. Fortunately, LOFAR has the capability to resolve the flux enhancements in the harmonic emission and such observations demonstrate
that the fluxes are disturbed in both the fundamental and harmonic components. Harmonic radio emission
is likely to be produced over a larger volume than the fundamental emission \cite{1987SoPh..111...89M}, hiding the spatial inhomogeneity of Langmuir waves. Indeed, LOFAR imaging of the harmonic
emission \cite{2017arXiv170806505K} shows that the harmonic component has larger source sizes than the fundamental emission.
Another possible reason is that the level of turbulence
is not strong enough to modulate the harmonic components.
From the simulation by \cite{2014ApJ...790...67L}, larger turbulence levels producing trough-peak regions in the plasma density profile may lead to broader, resolvable intensifications in the harmonic radiation,
while moderate turbulence levels yield flux enhancements with much broader half-power bandwidths in the fundamental emission,
which may account for the type IIIb-III pairs.
However, we have not found any way to estimate the levels of turbulence using present observations.

Remote and in situ spacecraft measurements have shown
that density perturbations are ubiquitous in the corona and solar wind,
and that they often exhibit a Kolmogorov power spectrum \citep[e.g.][]{1981Natur.291..561A,1987JGR....92..282M,1988PhFl...31.3634M,1995ARA&A..33..283G}.
The sensitive observations of solar radio radiation may provide
new insight into the properties of turbulent density perturbations
in the solar corona.
Here we used radio flux as a probe to estimate the properties of the background density perturbations. The spectral index of the density power-spectrum is found to be consistent with a Kolmogorov-like spectral index. The coronal turbulence may play an important role in the fine structure formation of solar radio type IIIb bursts and high resolution observations with LOFAR can provide new diagnostics to probe density fluctuations in the solar corona.

\acknowledgments
The work is supported by NSFC grants 11433006, 11373039, 11573039 and 11661161015 and by a Science \& Technology Facilities Council
(STFC) consolidated grant ST/L000741/1.
The work has also benefited from the support by the Marie Curie PIRSES-GA- 2011-295272 RadioSun project and from an international team grant (http://www.issibern.ch/teams/lofar/) from ISSI Bern, Switzerland.
This paper is based on data obtained with the International LOFAR Telescope (ILT). LOFAR \citep{2013A&A...556A...2V} is the Low Frequency Array designed and constructed by ASTRON. It has facilities in several countries, that are owned by various parties (each with their own funding sources), and that are collectively operated by the ILT foundation under a joint scientific policy. Some data are courtesy of NASA SDO science teams. The authors are thankful to Natasha Jeffrey for the help with the manuscript.

\bibliographystyle{apj}
\bibliography{cy}

\end{document}